\documentclass[letterpaper,conference]{IEEEtran}

\usepackage{cite}
\usepackage[cmex10]{amsmath}
\usepackage{mathtools}
\usepackage{url}
\usepackage{xspace}
\usepackage{siunitx}
\usepackage{tkz-graph}
\usetikzlibrary{calc}

\usepackage{caption}
\usepackage{subcaption}
\usepackage{multirow}
\usepackage{multicol}
\usepackage{stackengine}
\usepackage{wrapfig}
\usepackage{textcase}
\usepackage{ifthen}
\usepackage{cases}
\usepackage{amsmath} 
\usepackage{amsfonts}
\usepackage{paralist}
\usepackage{layouts}
\usepackage{booktabs}
\usepackage{makecell}
\usepackage{xspace}
\usepackage{tabularx}
\usepackage{todonotes}

\newcommand{\fakeparagraph}[1]{\noindent\textbf{#1.}}
\newcommand{\vect}[1]{\boldsymbol{#1}}

\newcommand{\rssi}{\ensuremath{\mathit{RSSI}}\xspace}
\newcommand{\wrt}{w.r.t.\ }

\newcommand{\home}{\textsc{home}\xspace}
\newcommand{\office}{\textsc{office}\xspace}
\newcolumntype{R}{>{\raggedleft\arraybackslash}X}
\newboolean{SUBMIT}
\newboolean{authnotes}

\setboolean{SUBMIT}{true} 

\ifthenelse{\boolean{SUBMIT}}
{
\usepackage{nohyperref}
\setboolean{authnotes}{false}
}{
\usepackage{hyperref}
\setboolean{authnotes}{true}
}

\ifthenelse{\boolean{authnotes}}
{
\newcommand{\ram}[1]{\footnote{{\bf Ramona: #1}}}
\newcommand{\adis}[1]{\footnote{{\bf Indika: #1}}}
\newcommand{\sam}[1]{\footnote{{\bf Sameera: #1}}}
\newcommand{\dirk}[1]{\footnote{{\bf Dirk: #1}}}
\newcommand{\piyush}[1]{\footnote{{\bf Piyush: #1}}}
\usepackage{draftwatermark}
\SetWatermarkLightness{0.85}
}{
\newcommand{\ram}[1]{}
\newcommand{\adis}[1]{}
\newcommand{\sam}[1]{}
\newcommand{\dirk}[1]{}
\newcommand{\piyush}[1]{}
}

\begin{document}
\title{Modeling WiFi Traffic for White Space Prediction in Wireless Sensor Networks}

\author{ 
	\IEEEauthorblockN{Indika S. A. Dhanapala$^{1,*}$, Ramona Marfievici$^{1}$, Sameera Palipana$^{1}$, Piyush Agrawal$^{2}$, Dirk Pesch$^{1}$} 
	\IEEEauthorblockA{
		$^{1}$Nimbus Centre for Embedded Systems Research, Cork Institute of Technology, Cork, Ireland \\
		$^{2}$United Technologies Research Centre, Cork, Ireland \\ 
		$^{*}$Contact Author: \texttt{I.S.A.Dhanapala@mycit.ie}
	} 
}

\maketitle

\setlength\abovedisplayskip{5pt}
\setlength\belowdisplayskip{5pt}

\setlength{\textfloatsep}{4pt plus 1.0pt minus 1.0pt} 
\setlength{\floatsep}{0pt plus 0.0pt minus 0.0pt} 


\begin{abstract}
Cross Technology Interference (CTI) is a prevalent phenomenon in the 2.4~GHz unlicensed spectrum causing packet losses and increased channel contention. In particular, WiFi interference is a severe problem for low-power wireless networks as its presence causes a significant degradation of the overall performance. 
In this paper, we propose a proactive approach based on WiFi interference modelling for accurately predicting transmission opportunities for low-power wireless networks. We leverage statistical analysis of real-world WiFi traces to learn aggregated traffic characteristics in terms of Inter-Arrival Time (IAT) that, once captured into a specific 2nd order Markov Modulated Poisson Process (MMPP(2)) model, enable accurate estimation of interference. We further use a hidden Markov model (HMM) for channel occupancy prediction. We evaluated the performance of \begin{inparaenum}[$i)$]
\item the MMPP(2) traffic model \wrt real-world traces and an existing Pareto model for accurately characterizing the WiFi traffic and,
\item compared the HMM based white space prediction to random channel access.
\end{inparaenum}
We report encouraging results for using interference modeling for white space prediction.
\end{abstract}

\IEEEpeerreviewmaketitle


\section{Introduction}
\label{sec:introduction}
Wireless technologies operating in unlicensed radio spectrum, such as the $2.4$~GHz ISM band, suffer from Cross Technology Interference (CTI), which is the time and frequency overlapping of concurrent transmissions. The interference occurs due to the broadcast nature of wireless transmissions of co-located devices whose radios are based on different technologies such as WiFi (IEEE802.11), Bluetooth (IEEE802.15.1) or IEEE802.15.4 and who cannot coordinate their transmissions. As a device cannot decode a data packet or MAC frame sent by a device of a different technology, which uses different modulation and coding schemes, CTI reduces the devices' ability to decode its own receiving signals. Consequently, CTI creates packet/frame losses, increases channel contention which increases delay, and ultimately under--utilizes the scarce frequency spectrum.

\begin{figure}[!hb]
	\centering
	\includegraphics[width=0.48\textwidth, height=0.49\textwidth, keepaspectratio]{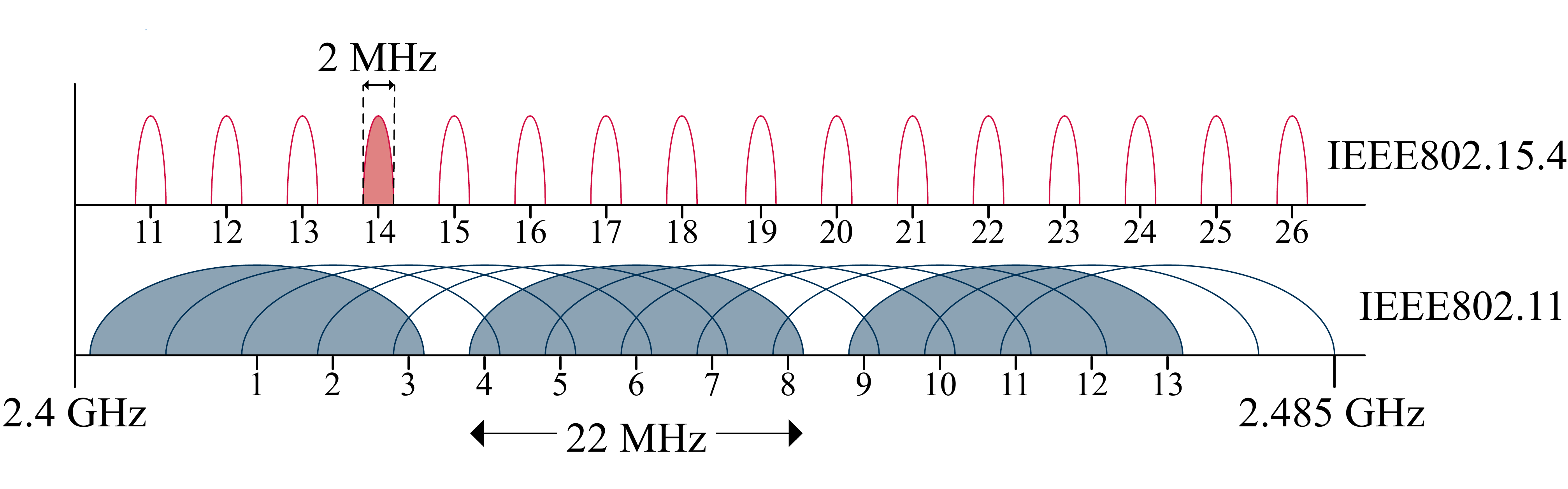}
	\caption{Overlapping WiFi and IEEE802.15.4 channels.}
	\label{fig:overlapping_channels}
\end{figure}

CTI caused by high power transmitters affects the lower power devices in particular, this is a problem for IEEE802.15.4 based WSNs in the $868$~MHz and $2.4$~GHz band as it reduces transmission opportunities and increases packet losses~\cite{LTE_alarm,Hithnawi2014}. 
WiFi is a particular problem as it dominates the $2.4$~GHz frequency band with much high transmission powers ($20$~dBm) than other technologies. In this study, we only consider coexistence of WiFi and IEEE802.15.4 based WSNs but the principles are fundamentally applicable to other technologies as well. 

As IEEE802.11 is a higher bandwidth technology than IEEE802.15.4, e.g. bandwidth ranges between $20$~MHz and $40$~MHz depending on the specific PHY layer specification, it creates interference for multiple $2$~MHz IEEE802.15.4 channels simultaneously (see Figure~\ref{fig:overlapping_channels}). This effect is called aggregated WiFi interference and reduces white spaces (the transmission free time periods between consecutive MAC frame transmissions) for WSNs, reduces transmission opportunities for WSNs and degrades WSN performance in terms of reliability and lifetime. Furthermore, the higher data rate of WiFi compared to IEEE802.15.4, e.g. $1$~Mbps to $300$~Mbps (depending on PHY layer) versus $250$~kbps creates rapid variations in time and frequency domain, which are hard to detect by the typical IEEE802.15.4 sensor node receiver hardware.

Existing solutions can be divided into reactive and proactive approaches to address CTI between WiFi and WSNs.
Reactive approaches include frequency domain solutions such as spreading and frequency hopping~\cite{conf/secon/IyerWL11}, the use of multiple antennae and transceivers to allow switching between different frequency bands when the current operating frequency band is interfered~\cite{kusy2011radio}, forward error correction schemes~\cite{Liang2010} and other time domain solutions such as CSMA/CA based on listen before talk (LBT)~\cite{boanobook}.
Proactive approaches try to predict white spaces, e.g. transmission opportunities for the low power technology when the high power technology is not transmitting~\cite{Huang2010,crosszig,boanojag}.
However, reactive approaches only deal with the problem when it occurs, thus negatively affecting performance and many approaches require changes in the respective standards, e.g. spreading codes, forward error correction or changes in the medium access protocol. On the other hand, proactive approaches can be used to quantify and predict when interference occurs if accurately modelled and can be used to predict when its best for the low power technology to transmit.

In this paper, we propose a proactive time domain technique to address CTI for WSNs. Our approach is based on \begin{inparaenum}[$i)$] \item an accurate WiFi traffic model to capture aggregated WiFi interference, and \item a model-based
technique to predict white spaces for IEEE802.15.4 transmissions.\end{inparaenum} The approach is motivated by our findings that WiFi traffic exhibits self--similarity at different time scales, which we found when examining real-world WiFi traffic traces and their packet Inter--Arrival Times (IAT) at different time resolutions (see Figure~\ref{fig:self_similarity}). 

We selected a 2{nd} order Markov Modulated Poisson Process (MMPP(2)) to model the WiFi packet IAT and a Hidden Markov Model (HMM) for predicting the white spaces. Our motivation for using an MMPP(2) model is that we need to model WiFi traffic for both saturated (i.e, busy) and unsaturated channel characteristics and we need to capture the self-similarity of WiFi traffic with only a few model parameters that are easily measurable. 

We learn WiFi aggregated traffic characteristics in terms of WiFi packet IAT from real-world WiFi traces and use those to calibrate the parameters of the MMPP(2) model, which provides us with an accurate model of aggregated WiFi interference. We then use the HMM for channel occupancy prediction, which is trained using the real-world WiFi traces and the MMPP(2) model. This provides us with a prediction when the channel is interference free and available to access.
We have evaluated the performance of the MMPP(2) traffic model with respect to the traces we collected and an existing Pareto model~\cite{Huang2010} and compared our HMM based channel access with 0.5-persistent random access technique.
Our analysis shows we can successfully predict the presence of white spaces. 

The rest of the paper is organised as follows. We provide a summary of existing techniques for interference detection and modelling for IEEE802.15.4 based WSNs in Section~\ref{sec:related_work}. Then, we discuss important properties of WiFi traffic and introduce our approach in Section~\ref{sec:proposed_technique}. The performance evaluation of the approach is presented in Section ~\ref{sec:perf_eval}. We discuss limitations of our work in Section~\ref{sec:discussion}, before the brief concluding remarks of Section~\ref{sec:conclusions}. 

\section{Related Work}
\label{sec:related_work}

\fakeparagraph{Detecting and classifying interference} 
Several works aim to measure and understand the impact of interference on WSNs, and classify interfering sources~\cite{Musaloiu-E2008b,Noda2011,Iyer2015,TIIM,sonic,crosszig}.
Musaloiu and Terzis~\cite{Musaloiu-E2008b} used \rssi based features to quantify the interference of all 16 IEEE802.15.4 channels to select the least interfered channel. Noda et al.~\cite{Noda2011} compute the ratio of channel idle time and channel busy time for assessing channel quality in the presence of interference. SpeckSense~\cite{Iyer2015} classifies \rssi bursts to characterize the channel as periodic, bursty or a combination of both. SoNIC~\cite{sonic} uses information from corrupted packets, therefore not incurring additional energy cost for \rssi sampling like the previous methods, to classify the sources of interference.  
However, these works succeed in detecting and identifying interference but it is not clear how these techniques are useful for autonomous interference mitigation due to the diversity of interferers. As a solution, TIIM~\cite{TIIM} extracts features from corrupted packets to quantify the interference conditions instead of identifying the interferer, thereby, the interference condition can be mapped to a specific mitigation technique. Nonetheless, TIIM does not present an implementation of these mitigation techniques; only in their follow up work, CrossZig~\cite{crosszig}, the authors have implemented an adaptive packet recovery and an adaptive FEC coding for this. 
All these solutions, however, are reactive, depending on the prevailing channel conditions, and do not aim to predict the white spaces through modeling,
which is instead our goal in this paper.

\fakeparagraph{Modeling the interference} 
Creating lightweight models of interference is not a trivial task. Several researchers have proposed models for channel occupancy~\cite{Stabellini2010,Huang2010,boanojag} or models for emulating interference caused by WiFi and Bluetooth~\cite{jamlab}. 
Stabellini and Zander~\cite{Stabellini2010} consider a general inter network coexistence scenario from a receiver centric perspective.
They propose a two--state semi--Markov model, in which, at a given time instant, a channel is identified as \textit{Busy} if a device transmits or \textit{Free} otherwise. They exploit the model for each node to identify the less interfered channel and to switch accordingly. 
In comparison, we consider the WiFi traffic perspective, and not that of a sensor receiver centric approach, and identify only the available white spaces of a specific channel through modeling WiFi traffic in time domain.
Boano et al.~\cite{Boano:2010:MSM:2127940.2127963,boanojag} define two--state semi--Markov model for channel occupancy and use noise measurements to measure the duration of the \textit{Free} and \textit{Busy} instants, and compute the CDF of \textit{Free} and \textit{Busy} periods. Based on the longest \textit{Busy} period, the authors derive MAC protocols parameters such that the application requirements are met.
JamLab~\cite{jamlab} models and regenerates interference patterns of different sources such as WiFi, Bluetooth and microwave ovens using sensor nodes. For modeling WiFi traffic, the authors considered both saturated (always \textit{Busy}) and unsaturated traffic scenarios. Saturated traffic is modeled with a Markov Chain model, while 
for the non--saturated WiFi a probability mass function of empirical data is used. In contrast, our goal is not to emulate WiFi traffic but to estimate it, and for this we use a single Markov Modulated Poisson Process (MMPP) to capture the existing traffic conditions, both saturated and non--saturated. 
The work presented in ~\cite{Huang2010}, instead, is closely related to our work since it focuses on a model--based white space prediction mechanism for WSNs in the presence of WiFi interference. Nevertheless, the model technique they used, Pareto, is a heavy--tailed distribution which usually captures long--range dependent (LRD) traffic behavior. However, our collected WiFi traces indicate that WiFi traffic is not exhibiting long--range dependency in the long run. 
Therefore, we use an MMPP model as it 
is not based on the LRD assumption. 

\section{Approach}
\label{sec:proposed_technique}
We provide the required background in traffic characterization in Section~\ref{ss:preliminaries} and then describe in details our approach. We begin by describing in Section~\ref{ss:overview_technique} the elements that are critical when using our approach for predicting white spaces. An integral element is the understanding of the WiFi traffic characteristics described in Section~\ref{ss:wifi_properties} and the actual modeling of the WiFi aggregated 
traffic, described in Section~\ref{ss:mmpp2}. Finally, in Section~\ref{ss:hmm} we introduce the HMM model supporting the prediction of white spaces.

\subsection{Preliminaries}
\label{ss:preliminaries}

\begin{figure}
	\centering
	\includegraphics[trim={0, 0.3cm, 0, 0.1cm}, clip, width=0.49\textwidth, height=0.49\textwidth, keepaspectratio]{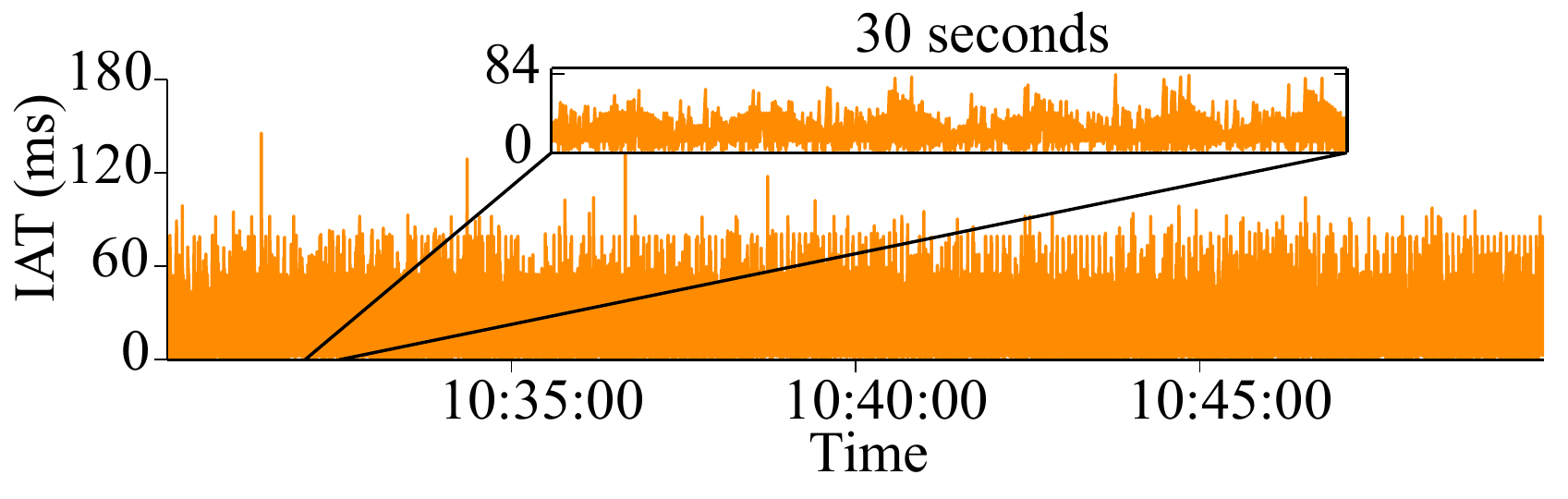}
	\caption[Self-similarity of WiFi aggregated traffic.]{Self-similarity of WiFi aggregated traffic. Statistics of 
	\begin{inparaenum}[a)] 
	\item full length: $M_{1}=18.85$~ms, $C=0.77$~ms, $H=0.52$,     
	\item zoomed part: $M_{1}=18.69$~ms, $C=0.79$~ms, $H=0.52$. 
	\end{inparaenum}}
	\label{fig:self_similarity}
\end{figure}

\fakeparagraph{Self--similarity}
A number of empirical studies of traffic measurements~\cite{Adas1997,Grossglauser1998,Muscariello20051835,Li2009} have convincingly demonstrated that the 
network traffic is self--similar. Intuitively, self--similarity describes the phenomenon in which certain characteristics (i.e., structural patterns and statistical properties) of the traffic are preserved irrespective of scaling in space or time, which can actually be exploited to better infer traffic properties.\\
To clarify our terminology, we briefly summarize the definition of a few basic concepts. Let $X(t) = (X_t: t \ge 0)$ be a stationary process, i.e., its joint distribution across a collection of times $t_1, \cdots, t_N$ is invariant to time shifting. $X$ is called \textit{self--similar} if: 
\begin{equation}
	X(at) = a^H X(t), a>0 
\end{equation}
where the equality refers to equality in distributions, $a$ is a scaling factor, and self--similarity parameter $H$ is called the \textit{Hurst} parameter. \\
The proposed approach takes advantage of this property,  depicted in Figure~\ref{fig:self_similarity} for a set of our traces, to infer traffic characteristics without incurring excessive data trace collection.

\fakeparagraph{Traffic statistical properties} 
Key statistical metrics typically used to provide insights on network traffic are: \textit{mean} ($\mu$), \textit{standard deviation} ($\sigma$) and, \textit{coefficient of variation} ($C$) computed as ${\sigma}$/${\mu}$. Moreover, the predominant way to quantify self--similarity is through \textit{H}, which is a scalar. \textit{H} takes on values from $0$ to $1$, with a value of $0.5$ indicating the absence of self--similarity, and the closer $H$ is to $1$, indicating the greater the self--similarity.
Calculating this parameter is not that straightforward: firstly, it can only be estimated, secondly, although there are several methods to estimate it, they often produce conflicting results. In our approach we used Peng, Periodogram and Box--Periodogram methods to estimate \textit{H} as they can produce a good estimate of \textit{H} for a sample size as low as $7000$~\cite{Rea2009}. The median of the three estimated \textit{Hurst} parameters is used, because the median is more stable than the mean even when one of the estimators fails. 
\subsection{Overview}
\label{ss:overview_technique}
\begin{figure*}[htp]
	\centering
	\includegraphics{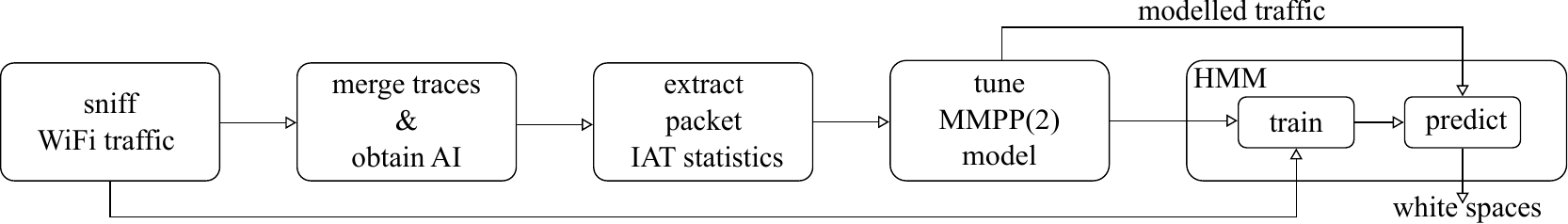}
	\caption{The system model.}
	\label{fig:system_model}
      \vspace{-6mm}
\end{figure*}
Figure~\ref{fig:system_model} illustrates the steps we follow for predicting white spaces for WSNs. To achieve our goal, we must start from assessing 
quantitatively the characteristics of the WiFi traffic in the operating environment. To this end, we collect a set of WiFi traces of length $x$~seconds and, in case of overlapping WiFi channels, we merge them to replicate the WiFi interference as seen by WSNs. Then, we extract the packet IAT distribution of the aggregated trace and characterize it in terms of mean~($M_1$), coefficient of variation~($C$) and \textit{Hurst} parameter~($H$). All the aforementioned statistical parameters are exploited to tune an MMPP(2) model for estimating the aggregated WiFi 
traffic for $y$~seconds. Then, we develop a Hidden Markov Model (HMM) with two states, \textit{Free} and \textit{Busy}, which is key for predicting transmission opportunities at run--time. The initial state probabilities are determined using the steady state probabilities of the MMPP(2) model, while the training is done using a set of WiFi traces of length $z$~seconds. 

\subsection{Determining WiFi Traffic Characteristics}
\label{ss:wifi_properties}

The approach we describe in this paper is based on an MMPP(2) model that:
\begin{inparaenum}[$i)$]
\item assumes traffic exhibits self-similarity and,
\item uses empirical data for estimating its parameters.
\end{inparaenum}
Therefore, we start by acquiring raw real--world WiFi traces from the operating environment and then process them in three steps:
\begin{enumerate}
\item \textit{aggregating}: it aggregates WiFi traces of multiple overlapping WiFi channels as seen by WSNs, i.e., concatenating raw traces and ordering them as a function of their timestamps. The output of this step is a trace of aggregated WiFi traffic;
\item \textit{traffic characteristics extraction}: the IAT distribution and its corresponding statistics in terms of mean~($M_1$), coefficient of variation~($C$) and \textit{Hurst} parameter~($H$) are determined from the aggregated WiFi traces. 
\item \textit{verifying self--similarity property}: the self--similarity property of the aggregated WiFi trace is determined by checking if the \textit{Hurst} parameter~($H$) satisfies the following condition: $H \in ($0.5$,$1$)$.
\end{enumerate}

\subsection{Estimating WiFi Traffic}
\label{ss:mmpp2}
We now describe how we exploit the processing just described towards building estimates of aggregated WiFi interference.
To this end we use an MMPP(2) model, as shown in Figure~\ref{fig:mmpp2}, whose defining parameters are (\cite{Fischer1993}):
\begin{equation}
\begin{split}
	\label{eq:inf_gen_rate}
  	\vect{Q} &= 
  	\begin{pmatrix}
    	-r_{1}  &  \phantom{-}r_{1} \\
    	 \phantom{-}r_{2}  & -r_{2}
  	\end{pmatrix} ,\quad
  \vect{\Lambda} = 
  \begin{pmatrix}
    \lambda_1 & 0           \\
    0 & \lambda_2
  \end{pmatrix}, \\
   \vect{\pi} &= \frac{1}{r_{1} + r_{2}}
	\begin{pmatrix}
	r_2 & r_1
	\end{pmatrix}	   
\end{split}
\end{equation}
where $\vect{Q}$ represents the infinitesimal generator, $\vect{\Lambda}$ is the matrix of the Poisson arrival rates, and $\vect{\pi}$ is the initial probability vector of the underlying Markov process. 
According to~\cite{Fischer1993,Nogueira1999}, an MMPP(2) process can be approximated by a second-order \textit{hyperexponential distribution}, with parameters $\mu_1$, $\mu_2$ and $p$, for fitting empirical packet IAT distribution to the model. As shown in Figure~\ref{fig:h2}, $\mu_1$, $\mu_2$ are mean packet arrival rates and $p$ represents the probability at which traffic is generated at a mean rate of $\mu_1$.

We exploit the output of the \textit{traffic characterization extraction}, computed mean time between arrival ($M_1$) and the coefficient of variation ($C$), to automatically compute the parameters $p$, $\mu_1$ and $\mu_2$ of the hyperexponential distribution, through the \textit{balanced means} method~\cite{Adan2015}: 
\begin{equation}
\begin{split}
p &= \frac{1}{2}\left( 1 + \sqrt{\frac{C^2 - 1}{C^2 + 1}}\right), \quad 
	\mu_1 = \frac{2p}{M_1}, \quad 
	\mu_2 = \frac{2(1-p)}{M_1}
    \end{split}
\end{equation}
Then, the parameters of the MMPP(2) model, $\lambda_1$, $\lambda_2$, $r_1$ and $r_2$ can be calculated from the estimated parameters $p$, $\mu_1$, $\mu_2$ and the Hurst parameter $H$, through: 
	\vspace{-2mm}
\begin{figure}[b]
\centering
\begin{subfigure}[t]{0.24\textwidth}
	\centering
	\includegraphics[trim={0, 0.6cm, 0, 0.5cm}, clip, width=\textwidth, height=0.5\textwidth, keepaspectratio]{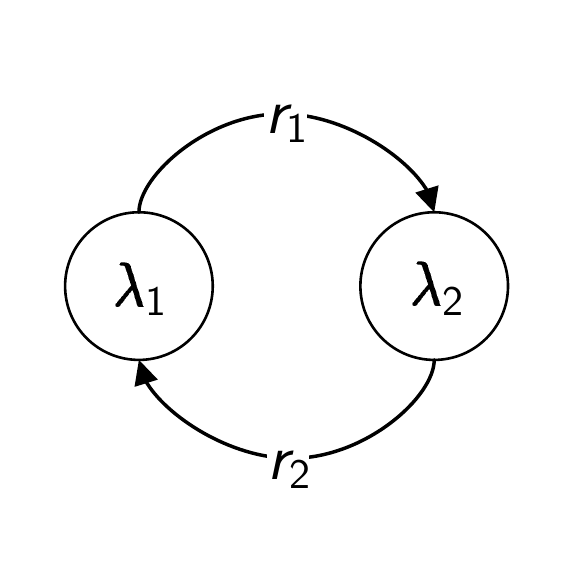}
	\caption{}
	\label{fig:mmpp2}
\end{subfigure}
\begin{subfigure}[t]{0.24\textwidth}
	\centering
	\includegraphics[trim={0, 0.6cm, 0, 0.5cm}, clip, width=\textwidth, height=\textwidth, keepaspectratio]{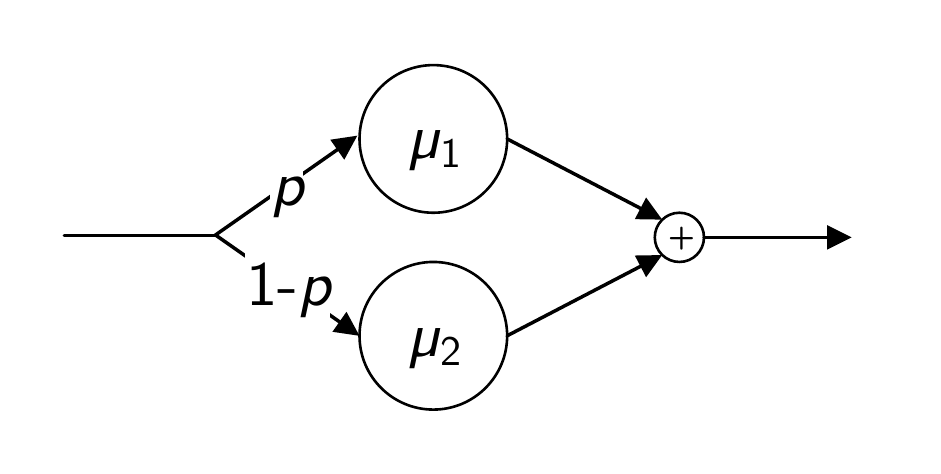}
    \caption{}
	\label{fig:h2}
\end{subfigure}
\caption{ (a) MMPP(2) and (b) $\mathrm{2}^{nd}$ order hyperexponential traffic models.}
\label{fig:mmpp2_h2}
\end{figure}
 
\begin{align}
\label{eq:mmpp2}
\begin{split}
	\lambda_1 &= \frac{1}{2} \left[ p(1-\beta)(\mu_1-\mu_2) + \beta\mu_1 + \mu_2 + \sqrt{\xi} \right] \\ 
	\lambda_2 &= \frac{\mu_1\mu_2[\lambda_1 - p(\mu_1 - \mu_2) - \mu_2]}{\lambda_1\mu_1 - \lambda_1p(\mu_1 - \mu_2) - \mu_1\mu_2} \\ 
	r_1 &= \frac{(\mu_1 - \lambda_1)(\mu_2 - \lambda_1)}{\lambda_2 - \lambda_1} \\ 
	r_2 &= \frac{(\lambda_2 - \mu_1)(\lambda_1 + r_1 - \mu_1)}{\mu_1 - \lambda_1} \\ 
\end{split}
\end{align}
where
\begin{align*}
	\xi &= \left[ p(1 - \beta)(\mu_1 - \mu_2) + \beta\mu_1 + \mu_2 \right]^2 - 4\beta\mu_1\mu_2, \\ 
	\beta &= 2 - 2H \\ 
\end{align*}
The approximation expressed through Eq.~\ref{eq:mmpp2} can be applied if and only if WiFi packet IAT distribution satisfies both conditions: $0.5 < H < 1$ and $C > 1$. In all the other cases, when $\frac{1}{\sqrt{2}} \leq C \leq 1$, a second--order \textit{Coxian distribution}~\cite{Adan2015} must be used prior to the second order \textit{hyperexponential distribution}. In this case, parameters $p$, $\mu_1$ and $\mu_2$ are computed, through:
\begin{align}
\label{eq:cox2}
\begin{split}
	p = \frac{1}{2C^2}, \quad 
	\mu_1 = \frac{2}{M_1}\left( \frac{p}{1+p}\right), \quad 
	\mu_2 = \frac{2}{M_1}
\end{split}
\end{align}
Note that $C<\frac{1}{\sqrt{2}}$ and $H<=0.5$ case was not considered in this work. 
This stage enables us to estimate and generate aggregated WiFi interference considering the particular profile of the observing environment. 

\subsection{Predicting White Spaces}
\label{ss:hmm}
As discussed in Section~\ref{sec:introduction}, the contribution we put forth here is an approach for predicting transmission opportunities for WSNs in the presence of WiFi interference. 
Next, we describe how we exploit the output of the previous stage, the estimated aggregated interference, along with real--world WiFi traces, to train an HMM that enables us to predict the wireless channel state. 
We adopt the notation presented in~\cite{Juang1986} to indicate the complete parameter set of the HMM: 
\begin{enumerate}
\item hidden (unobserved) states 
$\vect{S} = \{s_1, s_2\}$: correspond to two different regimes of the wireless channel, \textit{Free} and \textit{Busy};
\item initial state probabilities $\vect{\pi}$: in this context, they correspond to the steady state probabilities of the MMPP(2) model;
\item observations 
$\vect{O} = \{o_1, o_2\}$: correspond to two different values of the mean packet IAT of the MMPP(2) modeled trace, \textit{IAT\_small} and \textit{IAT\_large}. We define a threshold computed as the average over all samples' means for the training phase, and as a moving average of the mean WiFi packet IAT for the prediction phase. The choice of the moving average is motivated by the need to dynamically adapt the HMM to the operating environment. 
\item state transition probability matrix $\vect{A}$: models the evolution of the wireless channel as transitions among the set of unobserved states;
\item observation probability matrix $\vect{B}$.
\end{enumerate}
For a more in--depth discussion on how the observations are obtained and training procedure of the model, the reader shall refer to Section~\ref{ss:hmm_eval}. 
The model parameters $A$ and $B$ are properly initialized for the data set under consideration using uniformly distributed probabilities matrices and recomputed using the \textit{Baum–-Welch} algorithm~\cite{Juang1986}. 

\section{Performance Evaluation}
\label{sec:perf_eval}

We validate our approach for the white space prediction by:
\begin{inparaenum}[$i)$]
\item conducting a statistical comparison between empirical data traces (training set) we collected in indoor deployments, traces from the MMPP(2) model, and traces from a state-of-the-art Pareto model; 
\item comparing our predictions from HMM model with a 0.5--persistent random access method.
\end{inparaenum}
First, we present the experimental setup and how we acquired in-field WiFi traces. Then, we report and discuss our experimental results and show how to improve the accuracy by system calibration.

\subsection{WiFi Traces and Their Collection}
\fakeparagraph{Location}
In our study, we considered two indoor environments, \home and \office. This choice allowed us to exploit environments with different WiFi traffic saturations and to validate our approach under different conditions. While the \office has bursty traffic, it exhibits less self-similarity than the one in \home; confirmed by the values of \textit{H}: $0.5$ for \office and $0.7$ for \home.

\fakeparagraph{Hardware/software platform}
The WiFi dataset we use has been acquired by USB WiFi dongles that are 802.11n compliant. In both environments, the dongles were connected to a line-powered USB hub connected to a PC.
Because we are interested in exploiting IAT distributions, we created records of traces, i.e., \textit{pcap} files, of timestamped WiFi traffic using the \textit{tcpdump} tool.  
Then, a WiFi aggregated interference trace was obtained by merging traces of overlapping WiFi channels, i.e., traces from overlapping channels were appended and sorted in ascending order based on their timestamps. To this end, we used \textit{mergecap} tool and then \textit{tshark} to extract the IAT distribution. 

\fakeparagraph{Data collection execution}
To be useful, a model should accurately model traffic on any channel. In the \office environment we collected traces from four WiFi channels, i.e., $1$--$4$, ensuring each WSN overlapped channel, i.e., $11$--$14$, is affected by a different number of WiFi channels, as can be seen in Figure~\ref{fig:overlapping_channels}.
In the \home environment we wanted to collect traces under the same conditions as in \office but the environment did not exhibit that much WiFi traffic. Therefore, we resorted to sniffing WiFi channels $7$ to $13$ which overlap with WSN channels $20$ to $23$, each being affected by four WiFi channels. Moreover, to induce diversity in traffic on measured channels in \home environment we generated a continuous trace of video-streaming, using a laptop, on WiFi channel $11$ which only overlaps with WSN channels $21$, $22$, $23$.

The WiFi traces were collected during the day, for two hours, from 10:30AM to 12:30PM, in \office and \home. 

These data collection settings allow us to gather traces from channels in different environments that exhibit channel interference pattern diversity.

\subsection{System Calibration}
\label{ss:param_calibration}
\begin{figure}
	\centering
   	\includegraphics{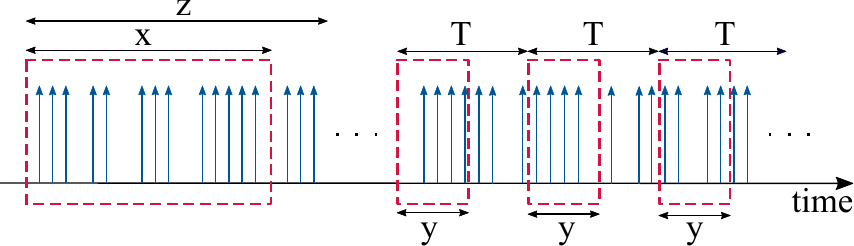}
	\caption{Approach timeline: $x$ is the MMPP(2) training duration, $y$ is the length of the modeled traffic, $z$ is the HMM training duration, while $T$ corresponds to the WSN traffic data rate.} 
	\label{fig:hmm_timeline}
\end{figure}

We start the evaluation by calibrating the training duration \textit{x}
and modelling duration \textit{y} of MMPP(2) model, and the training duration \textit{z} of the HMM model, defined as in Figure~\ref{fig:hmm_timeline}.

\fakeparagraph{MMPP(2) calibration}
The performance of the MMPP(2) model depends on parameters \textit{x} and \textit{y}.
For a low \textit{x}, the statistics from the training traces can not correctly capture the WiFi traffic behaviour, as the number of samples in such a small window is limited. Our logs indicate that a value of 
$x \leq 1$~second leads to this. A larger \textit{x} is equivalent with an increase in the length of the training trace which is prohibited in resource-constrained settings like the ones we target. 

Because we are interested in predicting the short-term behaviour of WiFi traffic, \textit{y} should be as small as possible. Moreover, in order to accurately model WiFi traffic,
the MMPP(2) model must generate traffic from both its states at least once. This leads to define a lower bound for \textit{y} denoted as $y_{lb}=\frac{1}{r_1}+\frac{1}{r_2}$, where $\frac{1}{r_i}$ represents the duration of a state in the MMPP(2) model. Therefore, during the calibration process, we used $y$ as an integer multiple of its lower bound $y_{lb}$ in order to find its optimum value ($y=y_{lb} \times k$, $k \in \mathbb{Z^+}$).

\begin{figure}
\begin{subfigure}[t]{0.24\textwidth}
	\centering
	\includegraphics[trim={0, 0.0cm, 0, 0.0cm}, clip, width=\textwidth, height=\textwidth, keepaspectratio]{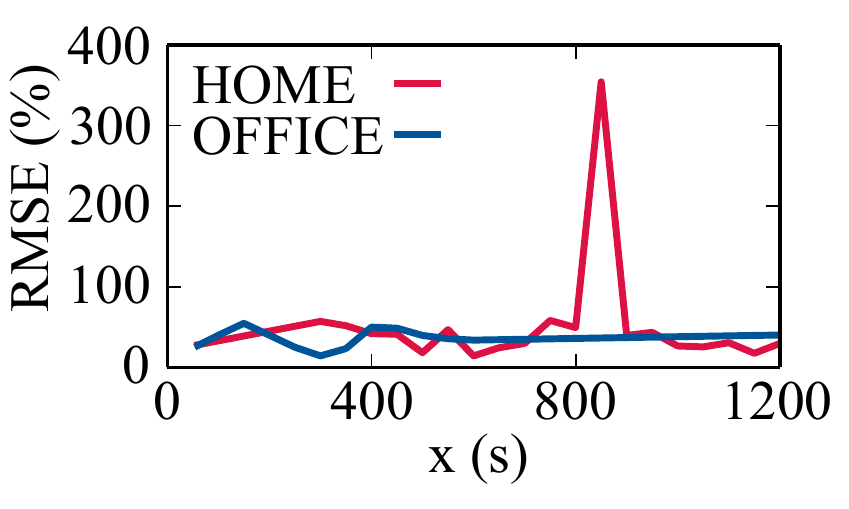}
\vspace{-3mm}
	\label{fig:rmse_x}	
\end{subfigure}
\begin{subfigure}[t]{0.24\textwidth}
	\centering
	\includegraphics[trim={0, 0.0cm, 0, 0.0cm}, clip, width=\textwidth, height=\textwidth, keepaspectratio]{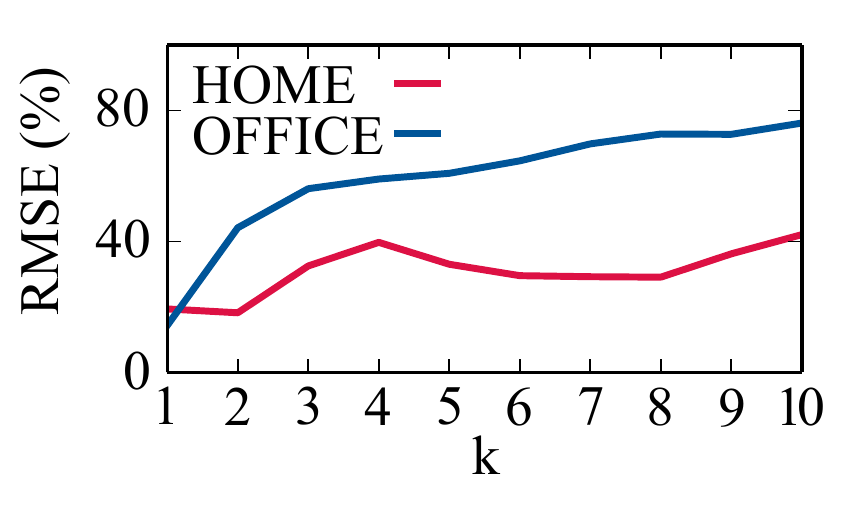}
    \vspace{-3mm}
	\label{fig:rmse_k}	
\end{subfigure}
\vspace{-6mm}
\caption{System calibration: MMPP(2) metrics variation with $x$ (left) and $k$ (right).} 
\label{fig:sys_calibration_mmpp2}
\end{figure}

As our goal is to reduce the RMSE between the modeled traffic and the testing set, we first played with different values for $x$ and $y$ to get a better understanding of their impact on the RMSE. 
Next, we fixed $y$ at $k=1$ in \office and $k=2$ in \home, and vary $x$ as an exponential function starting from $60$~seconds to $40$~minutes. 
Overall, all of our evaluations, as depicted in Figure~\ref{fig:sys_calibration_mmpp2}, show that a value of $1$ for $k$ and 300~seconds for $x$ and 2 for $k$ and 500~seconds for $x$ provide a minimum RMSE in \home and \office, respectively. 

\begin{figure}
\centering
   \begin{subfigure}[b]{0.5\textwidth}
   \includegraphics[width=1\linewidth, trim={0.7cm, 0.0cm, 0, 0.3cm}, clip, keepaspectratio]{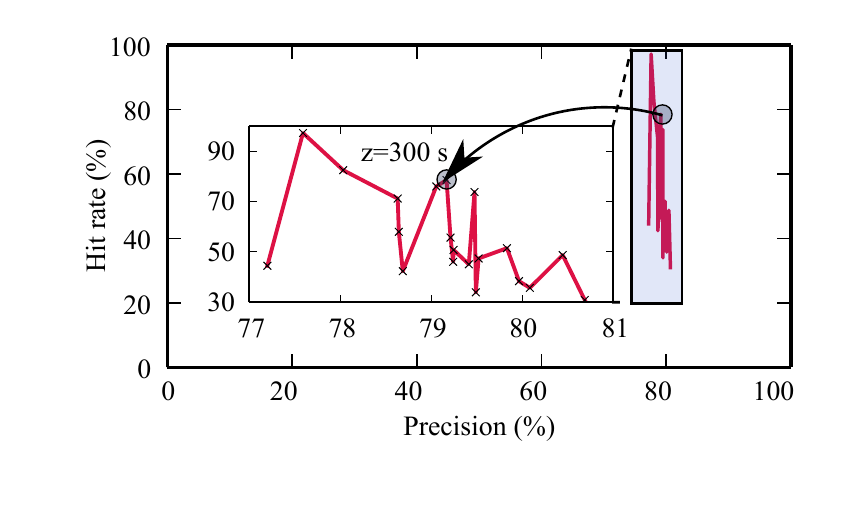}
   \vspace{-10mm}
   \label{fig:auc_office} 
\end{subfigure}
\begin{subfigure}[b]{0.5\textwidth}
   \includegraphics[width=1\linewidth, trim={0.7cm, 0.0cm, 0, 0.3cm}, clip, keepaspectratio]{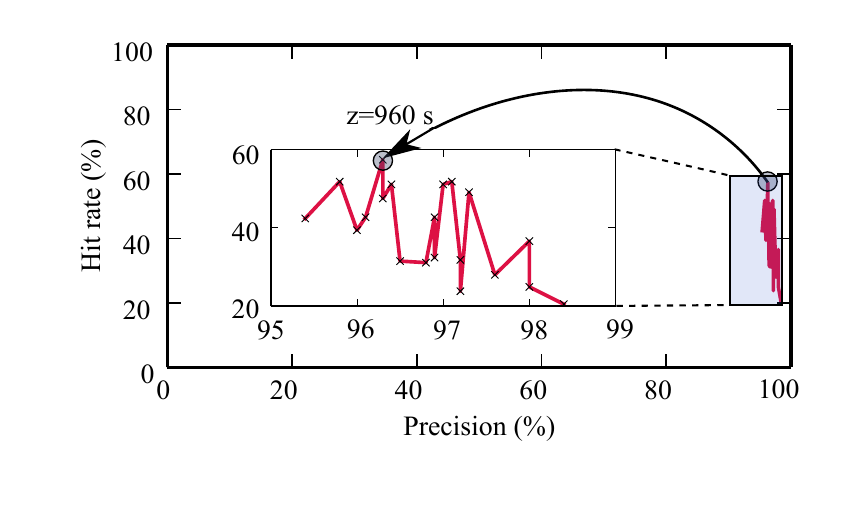}
    \vspace{-8mm}
   \label{fig:auc_home}
\end{subfigure}
\vspace{-8mm}
\caption{System calibration: HMM metrics variation with $z$ for \office(top) and \home(bottom).}
\label{fig:sys_calibration_hmm}
\end{figure}

\fakeparagraph{HMM calibration}
The performance of the HMM model depends on the size $z$ of the training set.
Our goal is to choose $z$ that maximizes the hit rate and minimizes the FDR. For this we used the precision-recall curve analysis~\cite{Bradley:1997:UAU:1746432.1746434}, see Figure~\ref{fig:sys_calibration_hmm}, looking at the impact of $z$ on the variation of hit rate and the precision (1-FDR) metric. Overall, all our evaluations show that a value of $300$~seconds for \office and $960$~seconds for \home satisfy the aforementioned criteria. Since we want to avoid the dependency between parameter $z$ and the WSN channel, we consider the average of hit rate and FDR over all channels.
 
\subsection{MMPP(2) Model Validation}
\label{ss:mmpp2_eval}

To evaluate the performance of our approach, we trained the MMPP(2) model with traces from the collected data traces from different environments and channels (i.e., length of training set: in \home for channels 21--23 is 6500~samples, for channel 20 is 3500~samples as it is not under the interference of the video-streaming traffic; in \office for channels 11--13 is 16000~samples, for channel 14 is 28000~samples as it is the channel affected by the highest number of WiFi interferers). We compare on the basis of RMSE statistics computed on the generated trace from the MMPP(2) model and a state-of-the-art Pareto model~\cite{Huang2010} versus an unseen trace (testing set) from the collected data traces.

Both models were trained with the same data trace, moreover, Pareto's scaling parameter was set to the maximum packet duration of a WSN packet. i.e., 4.256~ms, representing the minimum white space duration. 
\begin{figure}[t]
\centering
  \begin{subfigure}[t]{0.24\textwidth}
	\centering
	\includegraphics[trim={0, 0.0cm, 0, 0.0cm}, clip, width=\textwidth, height=\textwidth, keepaspectratio]{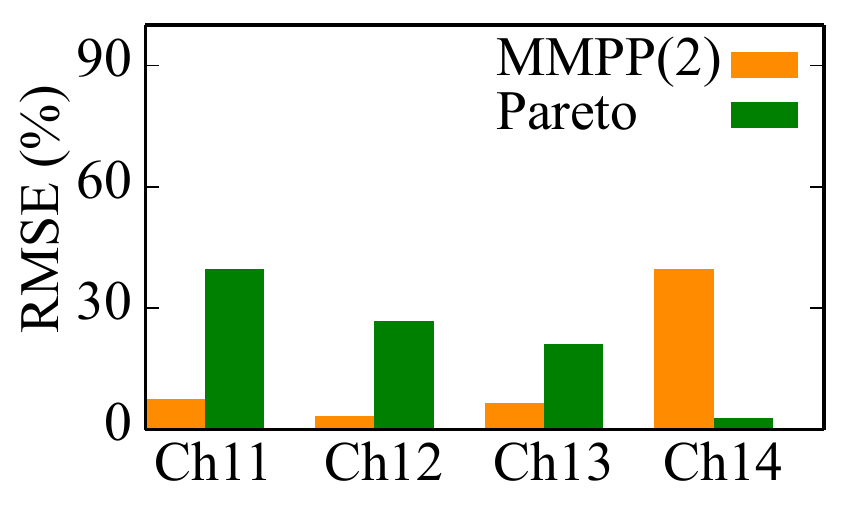}
    \vspace{-6mm}
	\label{fig:rmse_bar_office}
\end{subfigure}
\begin{subfigure}[t]{0.24\textwidth}
	\centering
	\includegraphics[trim={0, 0.0cm, 0, 0.0cm}, clip, width=\textwidth, height=\textwidth, keepaspectratio]{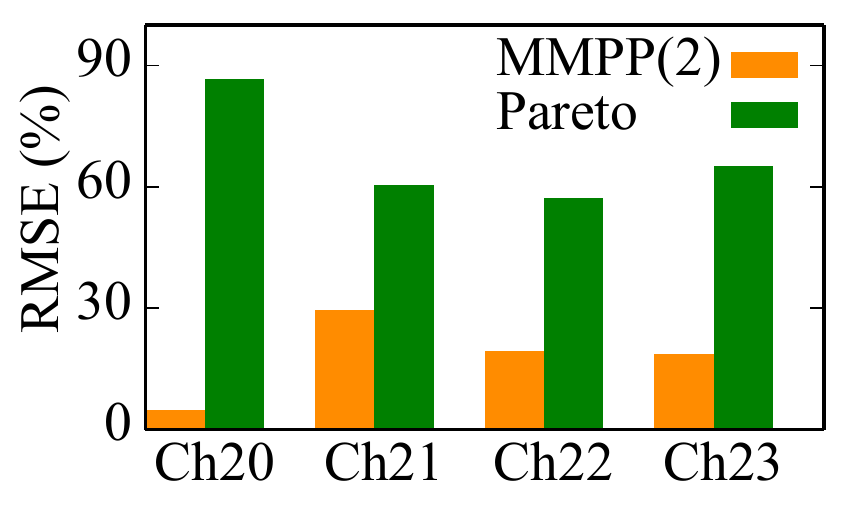}
    \vspace{-6mm}
	\label{fig:rmse_bar_home}
\end{subfigure}
\vspace{-5mm}
\caption{Comparison of RMSE: \office (left) and \home(right).} 
\label{fig:rmse_bar}
\end{figure}
Figure~\ref{fig:rmse_bar} shows the results from our comparison in terms of percentage RMSE. A few trends are clearly identifiable. First, the accuracy of the estimation decreases as one goes from \office to \home for both models--i.e., as the quantity of the interference decreases. However, this is not true for MMPP(2) on channel $14$ in \office and $20$ in \home.
Second, the accuracy of the estimation of MMPP(2) model is always higher than the Pareto model. The difference is more marked on channel 11 in \office and channel 20 in \home, Figure~\ref{fig:rmse_bar} showing a difference of $6.6$~ms and $244.5$~ms, respectively. However, channel $14$ in \office shows a different behaviour, Pareto delivering best performance on this channel even if there is more traffic. From Table~\ref{tab:traffic-char}, one can see that  
on channel $14$ there is a higher traffic, i.e., higher number of samples and lower mean IAT $10.7$~ms, than on the other channels, while the \textit{Hurst} parameter has the lowest value $0.51$, 
which translates to a less self-similar behaviour. The characteristics of the traffic on channel $14$ are more favourable for Pareto which captures the mean IAT than for MMPP(2) which requires all the three parameters for traffic modelling.
Moreover, on channel $20$, which is not overlapped by the video-streaming traffic channel, the mean of IAT is almost double, i.e., $141.5$~ms, than on other channels, i.e., $67$--$82$~ms, as can be seen in Table~\ref{tab:rmse_absolute}. As Pareto is a heavy-tailed distribution, it fails to capture the unsaturated traffic on this channel, having an RMSE of $259$~ms. On the other hand, MMPP(2) model performs the best, i.e, RMSE of $14.5$~ms out of all \home channels. 
All these results confirm the versatility of the MMPP(2) for modelling both unsaturated and saturated traffic. 

\begin{table}[t]
\centering
\caption{Traffic characteristics in \office and \home.}
\vspace{-2mm}
\label{tab:traffic-char}
\setlength\extrarowheight{2pt}
\begin{tabular}{|c|c|c|c|c|S[table-format=3.1]|c|c|}
\hline
\multicolumn{4}{|c|}{\textbf{\office}}                           & \multicolumn{4}{c|}{\textbf{\home}}                              \\ \hline
\textbf{\begin{tabular}[c]{@{}c@{}}WSN \\ Ch\end{tabular}} & \textbf{\begin{tabular}[c]{@{}c@{}}mean \\(ms)\end{tabular}} & \textbf{CV} & \textbf{Hurst} & \textbf{\begin{tabular}[c]{@{}c@{}}WSN \\ Ch\end{tabular}} & \textbf{\begin{tabular}[c]{@{}c@{}}mean \\(ms)\end{tabular}} & \textbf{CV} & \textbf{Hurst} \\ \hline
\textbf{11}      & 18.6          & 0.80        & 0.54           & \textbf{20}      & 141.5         & 0.90        & 0.63           \\ \hline
\textbf{12}      & 18.4          & 0.80        & 0.52           & \textbf{21}      & 67.1          & 0.92        & 0.69           \\ \hline
\textbf{13}      & 18.2          & 0.81        & 0.55           & \textbf{22}      & 76.0          & 0.91        & 0.70           \\ \hline
\textbf{14}      & 10.7          & 0.87        & 0.51           & \textbf{23}      & 82.2          & 0.92        & 0.71           \\ \hline
\end{tabular}
\end{table}

\begin{table}[]
\centering
\vspace{2mm}
\caption{Comparison of absolute RMSE.}
\vspace{-2mm}
\label{tab:rmse_absolute}
\setlength\extrarowheight{2pt}
\begin{tabular}{|c|c|c|c|c|c|}
\hline
\multicolumn{3}{|c|}{\textbf{\office}}                                                                                & \multicolumn{3}{c|}{\textbf{\home}}                                                                                   \\ \hline
\multirow{2}{*}{\textbf{\begin{tabular}[c]{@{}c@{}}WSN\\ Ch\end{tabular}}} & \multicolumn{2}{c|}{\textbf{RMSE (ms)}} & \multirow{2}{*}{\textbf{\begin{tabular}[c]{@{}c@{}}WSN\\ Ch\end{tabular}}} & \multicolumn{2}{c|}{\textbf{RMSE (ms)}} \\ \cline{2-3} \cline{5-6} 
                                                                           & \textbf{MMPP(2)}    & \textbf{Pareto}   &                                                                            & \textbf{MMPP(2)}    & \textbf{Pareto}   \\ \hline
\textbf{11}                                                                & 1.6                 & 8.1               & \textbf{20}                                                                & 14.5                & 259.0             \\ \hline
\textbf{12}                                                                & 0.7                 & 5.2               & \textbf{21}                                                                & 34.3                & 69.5              \\ \hline
\textbf{13}                                                                & 1.4                 & 4.4               & \textbf{22}                                                                & 21.4                & 62.9              \\ \hline
\textbf{14}                                                                & 4.2                 & 0.3               & \textbf{23}                                                                & 20.6                & 71.7              \\ \hline
\end{tabular}
\end{table}

\subsection{HMM Model Evaluation}
\label{ss:hmm_eval}

We now show that our HMM model approach provides accurate predictions of white spaces and compare it with a 0.5--persistent random access method (i.e., transmission attempt takes place with probability $0.5$).

We consider four metrics to assess the performance of our approach:
\begin{inparaenum}[i)]
\item \textit{hit rate},
\item \textit{false discovery rate (FDR)},
\item \textit{precision}, computed as $1-FDR$, 
and, 
\item \textit{F1 score}.
\end{inparaenum}

\begin{table}[b]
\centering
\captionsetup{width=.75\textwidth}
\caption{Confusion matrix.}
\vspace{-2mm}
\label{tab:cm}
\setlength\extrarowheight{2pt}
\begin{tabular}{c|c|c|c}
\cline{2-3}
                                              & \multicolumn{2}{c|}{\textbf{Predicted}} &                           \\ \hline
\multicolumn{1}{|c|}{\multirow{2}{*}{\textbf{Actual}}} & True Positives (TP)             & False Negatives (FN)            & \multicolumn{1}{c|}{\textit{Free}} \\ \cline{2-4} 
\multicolumn{1}{|c|}{}                        & False Positives (FP)             & True Negatives (TN)            & \multicolumn{1}{c|}{\textit{Busy}} \\ \hline
                                              & \textit{Free}           & \textit{Busy}          &                           \\ \cline{2-3}
\end{tabular}
\end{table}

\begin{table*}[!t]
\centering
\caption[bla]{Performance metrics of 0.5-persistent random access method and proposed HMM approach in two environments:
\begin{inparaenum}[a)]
\item \office: $x=300$~s, $y=y_{lb}$ and $z=300$~s, 
\item \home: $x=500$~s, $y=2 \times y_{lb}$ and $z=960$~s.
\end{inparaenum}}
\vspace{-2mm}
\label{tab:hmm_perf}
\setlength\extrarowheight{2pt}
\resizebox{\textwidth}{!}{%
\begin{tabular}{|c|c|c|c|c|c|c|c||c||c|c|c|c|c|c|l|l|}
\hline
\multicolumn{8}{|c}{\textbf{OFFICE}} & \multicolumn{1}{l}{} & \multicolumn{8}{c|}{\textbf{HOME}} \\ \hline
\textbf{\begin{tabular}[c]{@{}c@{}}WSN\\ ch\end{tabular}} & \textbf{\begin{tabular}[c]{@{}c@{}}FDR \\ (\%)\end{tabular}} & \textbf{\begin{tabular}[c]{@{}c@{}}Hit rate\\ (\%)\end{tabular}} & \textbf{\begin{tabular}[c]{@{}c@{}}F1 score\\ (\%)\end{tabular}} & \multicolumn{1}{l|}{\textbf{TP}} & \multicolumn{1}{l|}{\textbf{FP}} & \multicolumn{1}{l|}{\textbf{FN}} & \multicolumn{1}{l||}{\textbf{TN}} & \multicolumn{1}{l||}{\textbf{Method}} & \textbf{\begin{tabular}[c]{@{}c@{}}WSN\\ ch\end{tabular}} & \textbf{\begin{tabular}[c]{@{}c@{}}FDR\\ (\%)\end{tabular}} & \textbf{\begin{tabular}[c]{@{}c@{}}Hit rate\\ (\%)\end{tabular}} & \textbf{\begin{tabular}[c]{@{}c@{}}F1 score\\ (\%)\end{tabular}} & \textbf{TP} & \textbf{FP} & \textbf{FN} & \textbf{TN} \\ \hline
11 & \begin{tabular}[c]{@{}c@{}}22.2\\ 21.6\end{tabular} & \begin{tabular}[c]{@{}c@{}}50.6\\ 52.2\end{tabular} & \begin{tabular}[c]{@{}c@{}}61.3\\ 62.7\end{tabular} & \multicolumn{1}{c|}{\begin{tabular}[c]{@{}l@{}}543\\ 560\end{tabular}} & \multicolumn{1}{l|}{\begin{tabular}[c]{@{}l@{}}155\\ 154\end{tabular}} & \multicolumn{1}{l|}{\begin{tabular}[c]{@{}l@{}}530\\ 513\end{tabular}} & \multicolumn{1}{l||}{\begin{tabular}[c]{@{}l@{}}152\\ 153\end{tabular}} & \begin{tabular}[c]{@{}c@{}}random\\ HMM\end{tabular} & 20 & \begin{tabular}[c]{@{}c@{}}2.7\\ 2.8\end{tabular} & \begin{tabular}[c]{@{}c@{}}48.2\\ 61.1\end{tabular} & \begin{tabular}[c]{@{}c@{}}64.5\\ 75.0\end{tabular} & \begin{tabular}[c]{@{}c@{}}248\\ 314\end{tabular} & \begin{tabular}[c]{@{}c@{}}7\\ 9\end{tabular} & \multicolumn{1}{c|}{\begin{tabular}[c]{@{}c@{}}266\\ 200\end{tabular}} & \multicolumn{1}{c|}{\begin{tabular}[c]{@{}c@{}}7\\ 5\end{tabular}} \\ \hline
12 & \multicolumn{1}{l|}{\begin{tabular}[c]{@{}l@{}}23.1\\ 23.0\end{tabular}} & \begin{tabular}[c]{@{}c@{}}49.6\\ 90.8\end{tabular} & \begin{tabular}[c]{@{}c@{}}60.3\\ 83.4\end{tabular} & \multicolumn{1}{c|}{\begin{tabular}[c]{@{}c@{}}530\\ 970\end{tabular}} & \begin{tabular}[c]{@{}c@{}}159\\ 289\end{tabular} & \begin{tabular}[c]{@{}c@{}}538\\ 98\end{tabular} & \begin{tabular}[c]{@{}c@{}}153\\ 23\end{tabular} & \begin{tabular}[c]{@{}c@{}}random\\ HMM\end{tabular} & 21 & \multicolumn{1}{c|}{\begin{tabular}[c]{@{}c@{}}3.2\\ 3.7\end{tabular}} & \begin{tabular}[c]{@{}c@{}}53.3\\ 56.1\end{tabular} & \begin{tabular}[c]{@{}c@{}}68.6\\ 70.9\end{tabular} & \multicolumn{1}{c|}{\begin{tabular}[c]{@{}l@{}}269\\ 284\end{tabular}} & \multicolumn{1}{c|}{\begin{tabular}[c]{@{}c@{}}9\\ 11\end{tabular}} & \begin{tabular}[c]{@{}l@{}}237\\ 222\end{tabular} & \multicolumn{1}{c|}{\begin{tabular}[c]{@{}l@{}}13\\ 11\end{tabular}}\\ \hline
13 & \begin{tabular}[c]{@{}c@{}}0.0\\ 0.1\end{tabular} & \begin{tabular}[c]{@{}c@{}}50.2\\ 100\end{tabular} & \begin{tabular}[c]{@{}c@{}}66.8\\ 100\end{tabular} & \begin{tabular}[c]{@{}c@{}}692\\ 1379\end{tabular} & \begin{tabular}[c]{@{}c@{}}0\\ 1\end{tabular} & \begin{tabular}[c]{@{}c@{}}687\\ 0\end{tabular} & \begin{tabular}[c]{@{}c@{}}1\\ 0\end{tabular} & \begin{tabular}[c]{@{}c@{}}random\\ HMM\end{tabular} & 22 & \begin{tabular}[c]{@{}c@{}}3.8\\ 4.3\end{tabular} & \begin{tabular}[c]{@{}c@{}}50.4\\ 56.7\end{tabular} & \begin{tabular}[c]{@{}c@{}}66.1\\ 71.2\end{tabular} & \begin{tabular}[c]{@{}c@{}}256\\ 288\end{tabular} & \begin{tabular}[c]{@{}c@{}}10\\ 13\end{tabular} & \multicolumn{1}{c|}{\begin{tabular}[c]{@{}c@{}}252\\ 220\end{tabular}} & \multicolumn{1}{c|}{\begin{tabular}[c]{@{}c@{}}10\\ 7\end{tabular}} \\ \hline
14 & \multicolumn{1}{c|}{\begin{tabular}[c]{@{}l@{}}36.1\\ 37.5\end{tabular}} & \begin{tabular}[c]{@{}c@{}}49.8\\ 71.1\end{tabular} & \begin{tabular}[c]{@{}c@{}}56.0\\ 66.5\end{tabular} & \begin{tabular}[c]{@{}c@{}}431\\ 615\end{tabular} & \begin{tabular}[c]{@{}c@{}}243\\ 369\end{tabular} & \begin{tabular}[c]{@{}c@{}}434\\ 250\end{tabular} & \begin{tabular}[c]{@{}c@{}}272\\ 146\end{tabular} & \begin{tabular}[c]{@{}c@{}}random\\ HMM\end{tabular} & 23 & \begin{tabular}[c]{@{}c@{}}4.8\\ 4.0\end{tabular} & \begin{tabular}[c]{@{}c@{}}51.0\\ 55.9\end{tabular} & \begin{tabular}[c]{@{}c@{}}66.4\\ 70.6\end{tabular} & \begin{tabular}[c]{@{}c@{}}260\\ 285\end{tabular} & \begin{tabular}[c]{@{}c@{}}13\\ 12\end{tabular} & \begin{tabular}[c]{@{}c@{}}250\\ 225\end{tabular} & \multicolumn{1}{c|}{\begin{tabular}[c]{@{}c@{}}5\\ 6\end{tabular}} \\ \hline
\end{tabular}%
}
\vspace{-4mm}
\end{table*}
All metrics are derived from the elements of the confusion matrix in Table~\ref{tab:cm}, as follows:
\begin{align}
\footnotesize 
\label{eq:cm_metrics}
\begin{split}
	&\mathrm{Hit~rate}	=	\frac{\mathrm{TP}}{\mathrm{TP}+\mathrm{FN}}, \quad
	\mathrm{FDR}		=	\frac{\mathrm{FP}}{\mathrm{TP}+\mathrm{FP}}, \\ 
	&\mathrm{F1~score}	=	\frac{2(1-\mathrm{FDR}) \times \mathrm{hit~rate}}{2(1-\mathrm{FDR})+\mathrm{hit~rate}}
\end{split}
\end{align}

The first metric accounts for the  ability of the model to predict the channel is \textit{Free} when it is really  \textit{Free}, while the  second provides a direct assessment of the probability at which the model incorrectly identifies the channel as \textit{Free} when it is \textit{Busy}. In fact, FDR is a measure of the packet loss rate of the WSN when the prediction mechanism is being used. A good prediction mechanism should maximize the first metric, i.e., the hit rate, while minimizing the second, i.e., the FDR. The third metric, F1 score is defined as the harmonic mean between hit rate and precision metrics, balancing the two. Therefore, both metrics have to be high in order for F1 score to be high and the prediction to be good.

For the HMM training, a set of training observations and initial state probabilities are required. The latter are obtained from the MMPP(2) model and are constituted by the steady state probabilities, while the former are obtained from the sniffed $z$~seconds length  WiFi traffic as depicted in Figure~\ref{fig:hmm_timeline}. This step, in addition, extracts samples of $y$~seconds length every $T$~seconds, where $T$ is equal to the WSN data rate and $y$ is the modelled traffic duration. Next, the mean IAT of each sample is computed and compared with the predefined threshold, as defined in Section~\ref{ss:hmm}, in order to obtain the training observations sequence. 

Then, both the state transition  and observation probability matrix are properly initialized for the data set under consideration using uniformly distributed probabilities matrices and recomputed using the Baum--Welch algorithm.  
Based on these values, the calibrated  $x$, $y$ and $z$ values, and considering a WSN data rate $T = 5$~seconds, the HMM model is used for predicting white spaces every $T$~seconds.
We evaluated the performance of our approach using the four metrics and by comparing it with a $0.5$--persistent random access method with no retransmissions as Pareto, previously used for comparison in Section~\ref{ss:mmpp2_eval}, can not be used in conjunction with an HMM.

The results are shown in Table~\ref{tab:hmm_perf}.
We begin our analysis with the FDR metric which, despite its simplicity, provides an indicator of how our approach performs in avoiding collisions, i.e., packet losses. The quantitative data in the leftmost part of Table~\ref{tab:hmm_perf} shows that for all combinations of environments and channels our approach performs as good as the $0.5$--persistent access method. Moreover, our approach consistently performs better in \office than \home, apart from channel $13$. The reason for this behaviour is the fact that in \office the WiFi traffic exhibits heavy bursts which reduces the availability of white spaces compared to \home, while increasing the probability of  incorrectly discovering channel is \textit{Free}. This behaviour is evident if one looks through the lens of the FPs values. On the other hand, a single FP is observed on channel $13$, and our approach always predicts this channel is \textit{Free}, succeeding every time except once.

Moreover, looking at the hit rate metric, one grasps quickly that our approach performs much better in correctly predicting that the channel is \textit{Free} than the random method. Differences are more marked in \office and can be as high as $50$\%, except channel 11. 
We conjecture that this is an effect of the HMM state transition matrix, although this aspect requires further, finer-grained investigation.

Finally, the F1 score confirms that our approach is more efficient than the $0.5$--persistent method in identifying available white spaces in both environments.

\section{Limitations and Discussion}
\label{sec:discussion}

Our work is a first step towards CTI mitigation through WiFi aggregated traffic modeling and white spaces prediction. 
We validated our approach on real-world WiFi traffic traces from two environments each exhibiting different traffic characteristics, and show that it achieves good performance in both settings.
Despite the encouraging results, future work is required to validate the approach in different scenarios, i.e., different times (day/night) and locations with distinct traffic characteristics. 
Despite the self-similar traffic behaviour, during the day, the statistics characterizing the traffic might change, triggering the need to calibrate the system parameters. These additional experiments are already on our research agenda, and will enable us to ascertain to what extent our approach can be generalized.
In its current implementation, system calibration requires user control. Future work could optimize on this aspect and implement an approach that automatically recomputes the parameters when needed.
Our approach requires both WiFi transceivers and a sensor node to be collocated for the latter to perceive the same interference as seen by WiFi. However, this is not a barrier with the emergence of new chips such as Qualcomm~\cite{qualcomm} where multiple radio transceivers are collocated, i.e., WiFi, Bluetooth, 802.15.4.
Finally, a practical use of our technique would be its integration with both technologies.

\section{Conclusions}
\label{sec:conclusions}

Cross technology interference has been a crucial problem in wireless communication notably for devices operating in the unlicensed bands. While most existing solutions focus on detecting and classifying interference, less work has been done on interference modeling specially for mitigation purpose. In this work, we proposed a solution for accurately modeling WiFi aggregated traffic and predicting the presence of interference on the channel. We validated our WiFi traffic model against real--traces and a state--of--the--art Pareto model, and evaluated the prediction mechanism against 0.5--persistent random access method, both for saturated and unsaturated traffic. Results show that we are better in both settings. 
\vspace{5mm}

\fakeparagraph{Acknowledgements} We would like to thank Renato Lo Cigno from University of Trento, Italy, for his insightful comments on modelling WiFi traffic. This work has been funded by the Irish Research Council in collaboration with United Technologies Research Centre, Cork, Ireland. 

\bibliographystyle{IEEEtran}
\bibliography{References}



\end{document}